%% file: paper_arxiv.tex
\documentclass[10pt]{sig-alternate-05-2015}
\usepackage{graphicx}
\usepackage{subfig}
\usepackage{flushend}

\usepackage{xcolor}

\usepackage{blindtext}

\usepackage{etoolbox}
\makeatletter
\patchcmd{\maketitle}{\@copyrightspace}{}{}{}
\makeatother

\begin{document}





%

\title{Optimization Models for Flexible and Adaptive SDN Network Virtualization Layers}
	
%
%
%
%
%

\numberofauthors{1} 
%
\author{
%
%
Johannes Zerwas, Andreas Blenk, Wolfgang Kellerer\\
       \affaddr{Chair of Communication Networks, Technical University of Munich, Germany}\\
       \email{\{johannes.zerwas, andreas.blenk, wolfgang.kellerer\}@tum.de}
}

\maketitle

\input{abstract}

%
%

%
%

%
%


\input{introduction}
\input{model}
\input{evaluation}
\input{conclusion}

%
\bibliographystyle{abbrv}
\bibliography{2016hppjournal,bibliography}  
%
\end{document}

%% file: abstract.tex
\begin{abstract}
Network hypervisors provide the network virtualization layer for Software Defined Networking (SDN).
They enable virtual network (VN) tenants to bring their SDN controllers to program their logical networks individually according to their demands.
In order to make use of the high flexibility of virtual SDN networks and to provide high performance, the deployment of the virtualization layer needs to adapt to changing VN demands.
This paper initializes the study of the optimization of dynamic SDN network virtualization layers.
Based on the definition of reconfiguration events, we formalized mixed integer programs to analyze the multi-objective problem of adapting virtualization layers.
Our initial simulation results demonstrate Pareto frontiers of conflicting objectives, namely control plane latency and hypervisor and control path reconfigurations.
\end{abstract}

%% file: introduction.tex
\section{Introduction}
SDN network hypervisors provide the ability to control virtual networks (VNs) via SDN-based controllers.
Hypervisors connect tenant SDN controllers with their logical VNs.
Combining SDN and network virtualization provides the benefits of both worlds, i.e., flexible network resource sharing due to virtualization~\cite{Webb2011,Nikaein2015,Anderson2005} and programmability and adaptability due to open interfaces provided by SDN~\cite{McKeown2008,Sivaraman2013,Kreutz2015a}.
Accordingly, many network hypervisor solutions have already been proposed~\cite{Blenk2015a, Al-shabibi2014b,Sherwood2009, Jin2015,Rexford2014,Al-Shabibi2014,Bozakov2014,Bozakov2012}.
While these works only focus on the implementation aspect of network hypervisors, we draw attention to the occurring optimization problems when virtualizing SDN networks with focus on VN dynamics, e.g., due to changing user demands.

As it has been stated in the controller placement problem (CPP)~\cite{Heller2012}, the location of the control logic and the number of controllers can severely impact the performance of SDN networks.
For instance, long flow setup times can directly affect the perceived user quality, e.g., in case of web page requests where every millisecond counts~\cite{Wang2016b}.
Thus, as network hypervisors provide the control logic for virtualization, their placement inside the network needs to be optimized~\cite{Blenk2015hpp, Blenk2016}.
In order to pay attention to the dynamic nature of VNs~\cite{Keller}, e.g., due to changing user demands~\cite{Maier2009, Erman2013, Blenk2013}, we focus on the dynamic hypervisor placement problem (DHPP).

To always gain high efficiency, e.g., low control plane latency for VNs, hypervisor locations need to be adapted in case of changing VNs at runtime. 
However, such operation leads to reconfigurations of the existing hypervisor placement, e.g., migration of hypervisor entities or changing the routing of control paths between hypervisors and tenant SDN controllers.
Generally, network reconfigurations can lead to significant side effects, such as high latency or even network outages~\cite{Pathak2011,Gill2011}.
In this paper, we model the different reconfiguration types of virtual SDN networks.
Using mixed integer programming models for varying latency performance measures~\cite{Blenk2015hpp,Blenk2016}, we conducted an initial simulation study that focus on trade-offs between latency and reconfiguration measures.
Our initial results gain first insights into trade-offs among conflicting optimization goals, i.e., latency and reconfigurations.

%% file: model.tex
\section{Reconfiguration Models}
\begin{figure}
	\centering
	\subfloat[Initial]{
		\includegraphics[width=0.22\linewidth]{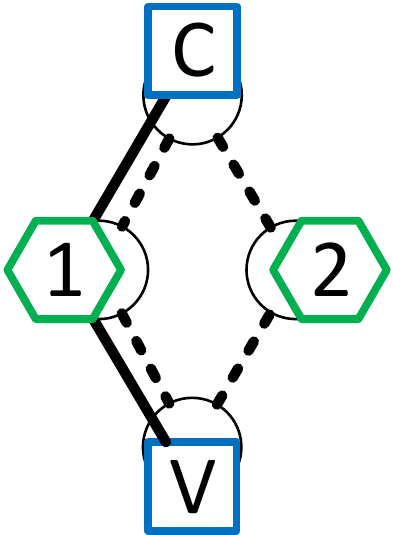}
		\label{fig:reconf_initial}
	}
	\subfloat[LOC]{
		\includegraphics[width=0.196\linewidth]{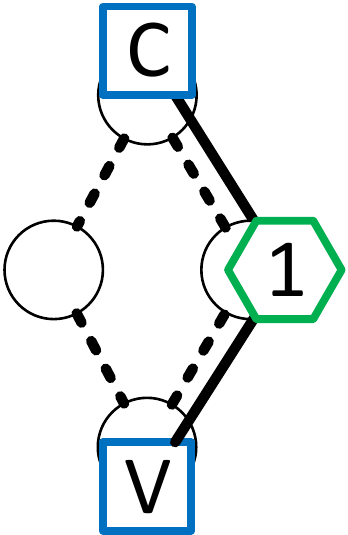}
		\label{fig:reconf_loc}
	}
	\subfloat[HV]{
		\includegraphics[width=0.22\linewidth]{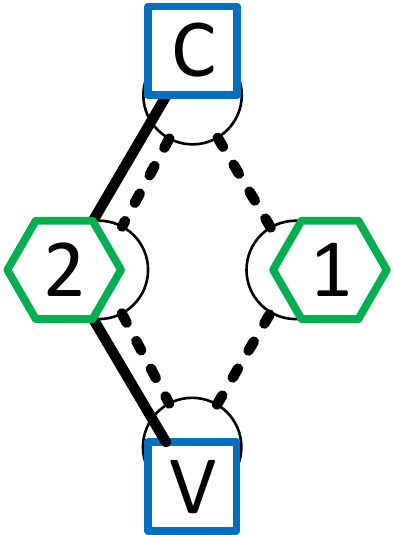}
		\label{fig:reconf_id}
	}
	\subfloat[Both]{
		\includegraphics[width=0.22\linewidth]{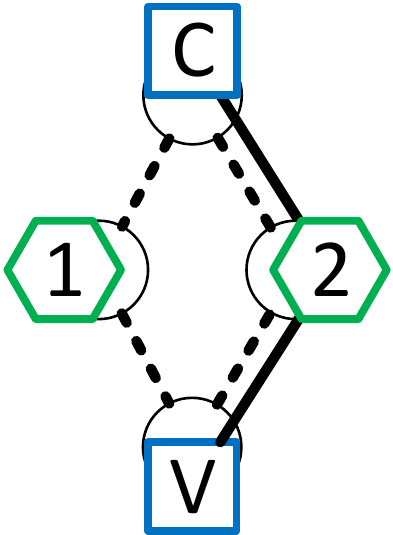}
		\label{fig:reconf_loc_id}
	}
	\hfill
	\caption{Reconfigurations for a virtual control path (VCP) in a virtualized SDN environment. Tenant controller (C), illustrated via boxes, connects via hypervisor (H), illustrated as hexagons, to virtual switch (C).
		}
	\label{fig:reconf}
\end{figure}
In virtual SDN environments, multiple tenant SDN controllers (SDN-C), hypervisor entities (HVs) and physical nodes hosting virtual switches (Vs) exist.
The SDN-Cs are connected via the hypervisor instances to their logical switches (Vs).
A virtual SDN control path (VCP) is identified via a tenant SDN controller (SDN-C), a logical SDN node (V), and a hypervisor entity (HV), which connects both.
Inside a network, a hypervisor entity (HV) is placed at a specific location (LOC).
The solution of a static hypervisor placement, i.e., one that does not consider reconfigurations, provides the location of  HVs and the routing of VCPs, i.e., which tenant controller is connected via which entity to its logical switch, for an objective like average control plane latency.

In case of adapting an HV placement, three different VCP reconfiguration events can be classified.
Fig.~\ref{fig:reconf} shows an initial HV placement (a) and the three reconfigurations (b-d) for one controller (C) and its logical switch (V) (both illustrated via squares), and two HV instances 1 and 2 (ill. via hexagons).
The substrate consists of four nodes (circles) and four physical links (dashed lines).
In the initial placement (Fig.~\ref{fig:reconf}(a)), the VCP (solid line) between V and C is routed via HV~1.
A hypervisor migration, i.e., VCP location change (LOC), is shown in Fig.~\ref{fig:reconf}(b) where HV~1 is migrated to the right node, whereas HV~2 is shutdown, e.g., due to energy savings.
Here, the VCP is still routed via HV~1, however the routing has been adapted.
Fig~\ref{fig:reconf}(c) shows an HV change for the VCP.
Here, the HV entity changed whereas the VCP routing stays.
Finally, the HV entity and the routing (LOC) are changed in (Fig.~\ref{fig:reconf}(d)).
For considering the reconfigurations while solving the HPP, two metrics provide the number of reconfigurations of VCPs between the current and the new HV placement.
$R_{VCP,LOC}$ determines the number of VCP location changes (Fig.1(a)). $R_{VCP,HV}$ determines VCPs that migrated from one HV entity to another.

%% file: evaluation.tex
\section{Initial Optimization Results}
\begin{figure}
	\centering
	\subfloat[LOC reconfigurations]{
		\includegraphics[width=0.49\linewidth]{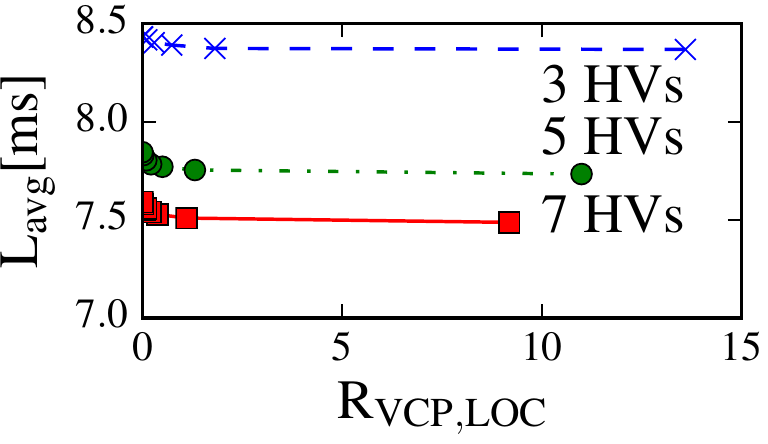}
	}
	\subfloat[HV reconfigurations]{
		\includegraphics[width=0.49\linewidth]{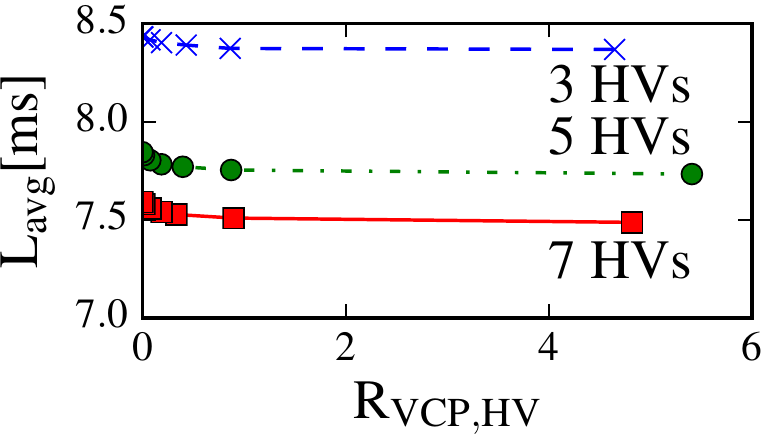}
	}\hfil
	\subfloat[LOC reconfigurations \newline ($5$ HVs)]{
		\includegraphics[width=0.49\linewidth]{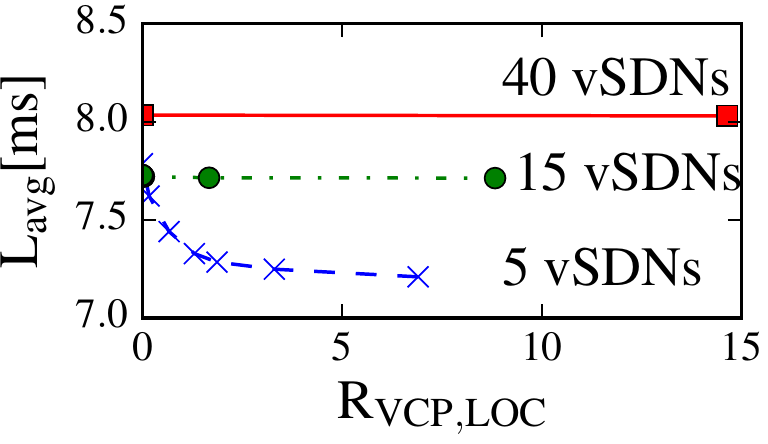}
	}
	\subfloat[HV reconfigurations \newline ($5$ HVs)]{
		\includegraphics[width=0.49\linewidth]{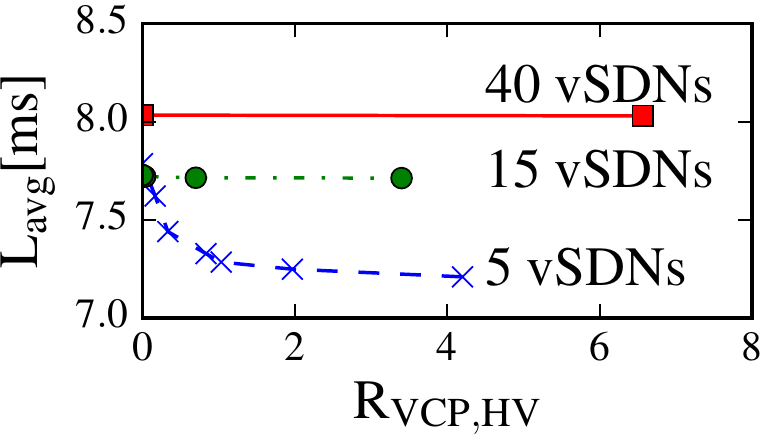}
	}\hfil
	\caption{Figures sowing trade off between average control plane latency $L_{avg}$ and number of reconfigurations.}
	\label{fig:pareto_front}
\end{figure}
To evaluate the interaction of the latency and reconfiguration objectives, our existing HPP framework \cite{Blenk2015hpp, Blenk2016} is extended to maintain initial HV placements (HP) and address the reconfiguration objectives.
We investigate the dynamic hypervisor placement problem for the case where an additional vSDN needs to be added to the network, i.e., for an increasing number of hosted vSDNs.
In particular, we would like to analyze the impact of adding vSDNs on the average control plane latency and the amount of reconfigurations, which are needed to adapt the HP towards an optimal latency placement.

Multi-stage optimization is considered as approach to perform multi-objective optimization for three objectives, namely minimum average HP control plane latency ($L_{avg}$), minimum LOC changes, and minimum HV entity changes.
In Stage~1, the average HP control plane latency ($L_{avg}$ in milliseconds [ms]) is optimized.
The LOCs are minimized in Stage~2.
In Stage~3, the HV entity changes are minimized.
The latency value obtained in Stage~1 is used to upper bound the latency of Stage~2 and 3. 
A latency relaxation factor $\rho$ in Stage~1 and 2 is introduced to determine the Pareto frontier.

The ATT MPLS network from the Topology Zoo~\cite{Knight2011} is used as physical network. 
The size and embedding of the virtual SDN networks (vSDNs) is randomly distributed on the physical network. The number of initially embedded vSDNs varies between 5 and 40.
For each number of embedded vSDNs, 30 different sets are generated to gain statistical confidence.
We analyze the impact of the number of HVs ($3,5,7$) and the impact of the number of vSDNs ($5,15,40$).
Fig.~\ref{fig:pareto_front}(a-b) show $L_{avg}$ against the number of reconfigurations for $3,5,7$ HVs and $5-40$ vSDNs. 
The marker represent the mean values over number of vSDNs and runs for $\rho$ varying between $0\,\%$ and $10\,\%$.
We observe that a small latency relaxation of $1\%$ already decreases both $R_{VCP,LOC}$ and $R_{VCP,HV}$ significantly.
For larger $\rho$, the reduction of the reconfiguration metrics diminishes.
A $\rho=\,10\%$ results in avoiding reconfigurations on average.

Fig.~\ref{fig:pareto_front}(c)-(d) illustrate the impact of initially embedded vSDNs on the reconfigurations for $5$ HVs.
For $5$ vSDNs, avoiding reconfigurations requires a larger latency increase than for setups that contain more vSDNs.  
A latency increase from $7.17\,\mathrm{ms}$ to $7.67\,\mathrm{ms}$ ($7\,\%$) is required to reduce $R_{VCP,LOC}$ to $0$.
In contrast, for $40$ vSDNs only a latency increase of about $0.017$ms avoids all reconfigurations.
Considering $R_{VCP,ID}$, similar observations are made.
Thus, the number of vSDNs should be taken into account while performing the DHPP.

%% file: conclusion.tex
\section{Conclusions}
SDN network hypervisors enable tenants to bring their own controller.
In this paper, we initialized the study of the dynamic hypervisor placement problem.
The dynamic problem introduces conflicting objectives, namely latency and reconfigurations.
Based on mixed integer programming formulations, we conducted a study of the multi-objective optimization problem.
We conclude that a latency relaxation of $7\%$ already leads to no reconfigurations at all.
Generally, our models can serve as baselines when designing heuristic solutions.
For future work, we plan to extend our study with different objectives and analyze it for larger substrate networks.

\section*{Acknowledgment}
This work has been performed in part in the framework of
the CELTIC EUREKA project SENDATE-PLANETS (Project
ID C2015/3-1) and is partly funded by the German BMBF
(Project ID 16KIS0473), and in part in the framework of the
EU project FlexNets funded by the European Research Council
under the European Unions Horizon 2020 research and
innovation program (grant agreement No 647158 - FlexNets).
The authors alone are responsible for the content of the paper.
